\documentclass[preprint]{sigplanconf}
\usepackage{xcolor}
\usepackage{amsmath}
\usepackage{graphicx}
\usepackage{setspace}
\usepackage{amssymb,epsfig,amstext,graphics}
\usepackage{times}
\usepackage{indentfirst}
\usepackage{wasysym}
\usepackage{booktabs}
\usepackage{stmaryrd}
\usepackage{mathrsfs}
\usepackage{multirow}
\usepackage{multicol}
\usepackage{amssymb}
\usepackage{amsfonts}
\usepackage{array}
\usepackage{booktabs}
\usepackage{threeparttable}

\newcommand{\term}[1]{\texttt{\small #1}}

\usepackage[colorlinks,
            linkcolor=black,
            anchorcolor=black,
            citecolor=blue,
			urlcolor=black,
            bookmarks=true
            ]{hyperref}
 \usepackage{listings}
          
\begin{document}

\setlength{\pdfpageheight}{\paperheight}
\setlength{\pdfpagewidth}{\paperwidth}

\conferenceinfo{CONF 'yy}{Month d--d, 20yy, City, ST, Country}
\copyrightyear{20yy}
\copyrightdata{978-1-nnnn-nnnn-n/yy/mm}
\copyrightdoi{nnnnnnn.nnnnnnn}

% Uncomment the publication rights you want to use.
%\publicationrights{transferred}
%\publicationrights{licensed}     % this is the default
%\publicationrights{author-pays}

\titlebanner{banner above paper title}        % These are ignored unless
\preprintfooter{short description of paper}   % 'preprint' option specified.

%\title{Towards a formal automobile RTOS standard in K framework for conformance verification}
%\subtitle{Subtitle Text, if any}
\title{Towards an executable semantics of automobile RTOS standard and its application to conformance verification}

%
%\author{\IEEEauthorblockN{Xiaoran Zhu\IEEEauthorrefmark{1}, Yuanmin Xu\IEEEauthorrefmark{2}, Jian Guo\IEEEauthorrefmark{1}$^{1}$, Xi Wu\IEEEauthorrefmark{2}, Huibiao Zhu\IEEEauthorrefmark{2}, Weikai Miao\IEEEauthorrefmark{2}}
%\IEEEauthorblockA{
%    \IEEEauthorrefmark{1}Soft/Hardware Co-design Engineering Research Center, East China Normal University\\
%    \IEEEauthorrefmark{2}Shanghai Key Laboratory of Trustworthy Computing, East China Normal University\\
%    {Email: \IEEEauthorrefmark{1}\IEEEauthorrefmark{2}\{sei\_xrzhu, sei\_ymxu\}@126.com, \IEEEauthorrefmark{1}\IEEEauthorrefmark{2}\{jguo, xiwu, hbzhu, wkmiao\}@sei.ecnu.edu.cn}}
%}
% 

\authorinfo{Xiaoran Zhu}
           {Soft/Hardware Co-design Engineering Research Center, East China Normal University}
           {zhuxrsandra@163.com}

\authorinfo{Min Zhang}
           {Shanghai Key Laboratory of Trustworthy Computing, East China Normal University}
           {zhangmin@sei.ecnu.edu.cn}
           
\authorinfo{Jian Guo}
           {Soft/Hardware Co-design Engineering Research Center, East China Normal University}
           {jguo@sei.ecnu.edu.cn}   

\maketitle

\begin{abstract}
The automobile Real-Time Operating System (RTOS) is hard to design and implement due to its real time features and increasing complexity. Some automobile RTOS standards are released aiming at unifying the software architecture of vehicle systems. Most of the standards are presented informally in natural languages, which may lead to not only ambiguities in specifications but also difficulties in conformance verification. This paper proposes a rewriting-based  approach for formalising the automobile RTOS standard. Taking the OSEK/VDX standard as an example, an executional formal semantics of the automobile RTOS kernel, which focuses on the real time features, is defined using $\mathbb{K}$, a rewriting-based framework. We also report some ambiguous definitions of the OSEK/VDX standard, which we find in the process of formalisation.  The $\mathbb{K}$ semantics of the OSEK/VDX standard is applied to conformance verification, which is used to check the conformance of not only the automobile operating system kernel but the applications.

% an industrial application for the verification of its conformance to the OSEK/VDX standard, which demonstrates the usefulness of the proposed approach.
 %We propose an approach called conformance verification to system analysis with K semantics and conduct a case study to it, which demonstrates the usefulness of the proposed method.

\end{abstract}

% general terms are not compulsory anymore,
% you may leave them out


\section{introduction}
\par Vast arrays of applications are increasingly used in automobile RTOSs to improve the functionality of the system. As a result, the automobile RTOS becomes more complex to design and implement. Some automobile RTOS standards have been put forward, with the purpose of improving the reusability of automobile applications and unifying the development process of software.
%\par RTOS (real time operating system) has been applied in many embedded systems due to its responsiveness and high reliability. The automobile RTOS is the core of vehicle system. Its performance has a big influence on the safety of vehicle. With the purpose of improving the reusability of automobile applications and unifying the development process of software, the researchers put forward some automobile RTOS specifications.
\par The automobile standards, such as OSEK/VDX \cite{OSEK}, one of the most widely used automobile RTOS standard, define standard interfaces of operating system, network management and so on. 
The automobile RTOS venders are required to achieve a certification through some technical methods, such as conformance testing \cite{conformancetesting}. The basic idea of conformance testing is to test the properties of automobile RTOSs with a test suite, which is generally derived from the standards. However, most of the standards are given in natural languages, which may cause ambiguities \cite{OSEKreview}. The ambiguity may mislead not only developers on the implementations but testing engineers on the composition of test cases. Moreover, test cases may not cover all possibilities of the test targets. All these facts lead to the result of conformance testing incomplete. Conformance verification \cite{con-verification} is another complementary approach, by which a specification in natural language is first formalised, and an implementation is verified whether it satisfies certain properties with respect to the formal specification.

%which verifies certain properties of an implementation with respect to its formal specification, is another complementary approach based on formal method \cite{formalmethod}. 

\par Conformance verification is a kind of so-called formal method \cite{formalmethod}, which is a mathematical technique for the specification, development and verification of computer systems. 
 For its accuracy and unambiguity, formal method has been widely applied in many areas, including embedded systems. IBM verified the power gates \cite{IBM1}, registers \cite{IBM2} and IBM Power7 microprocessor \cite{IBM3} formally to guarantee the reliability of their products. Intel Core i7 processor has been formally verified before being introduced to the market \cite{Intel}.
 Formal method has also been used to analyze and verify automobile RTOSs and applications. 
In \cite{SMT},  an approach was proposed to verify the applications of automobile operating systems by bounded model checking. In \cite{formal model,choi2013}, a formal model of an automobile RTOS is defined to generate test suite. In \cite{OSEK-K}, an executable formal semantics of a part of the OSEK/VDX-based automobile RTOS (OSEK/VDX RTOS for abbreviation) is defined using $\mathbb{K}$ framework \cite{K} for the model checking of user-defined automobile applications. However, a common weak point of the mentioned existing work is that they do not consider  the real time feature, which plays an important role in automobile RTOSs. 

\par 
By extending the work \cite{OSEK-K}, we propose in this paper a more complete formal semantics of the OSEK/VDX standard with considering its real time feature, and apply the formal semantics in conformance verification of an industrial automobile operating system. The formal semantics covers more features of the standard, such as the multiple activation of tasks and the alarm mechanism (see Section \ref{subsec:multi} and \ref{subsec:alarm}). In the process of formal analysis, we realise some ambiguous description in the OSEK/VDX standard, and give them formal definitions on the basis of our understandings and discussion with automobile electronical engineers. We show with a concrete case study how the conformance verification is achieved by the K semantics. In the case study, we detect some inconformities of an industrial operating system to the standards. After fixing the inconformities, we verified the conformance of the operating system by LTL model checking with respect to some important properties, such as starvation freedom.
\par In summary, the contributions of this paper are following:
\begin{enumerate}
\item An executable formal semantics of automobile RTOS kernel based on OSEK/VDX standard is given. The formal semantics, which is defined using $\mathbb{K}$ framework, not only specify the execution rules of OSEK/VDX system services, but also take the real time features into account. 
\item Some ambiguous descriptions in the OSEK/VDX standard are reported, and their corresponding formal semantics are defined. 
\item A new approach to system analysis with the K semantics called {\it{conformance verification}} is proposed. To demonstrate the usefulness of the approach, we conduct a case study on an industrial RTOS to it.
%\item A case study with an industrial application called EMS (Engine Management System) is conducted, which demonstrate the usefulness of the formal semantics to conformance verification.
\end{enumerate}
\par This paper is organized as follows. In Section 2, a brief introduction of the OSEK/VDX operating system is given. We analyze the ambiguous definitions of the OSEK/VDX standard in Section 3. Section 4 gives a formal description of the standard. Then an industrial application is analysed and verified in section 5. Section 6 and section 7 discuss some related work and conclude this paper respectively.

\section{Key Conceptions in the OSEK/VDX standard}

\begin{figure}[htbp]
\begin{flushleft}
\includegraphics[width=0.48\textwidth]{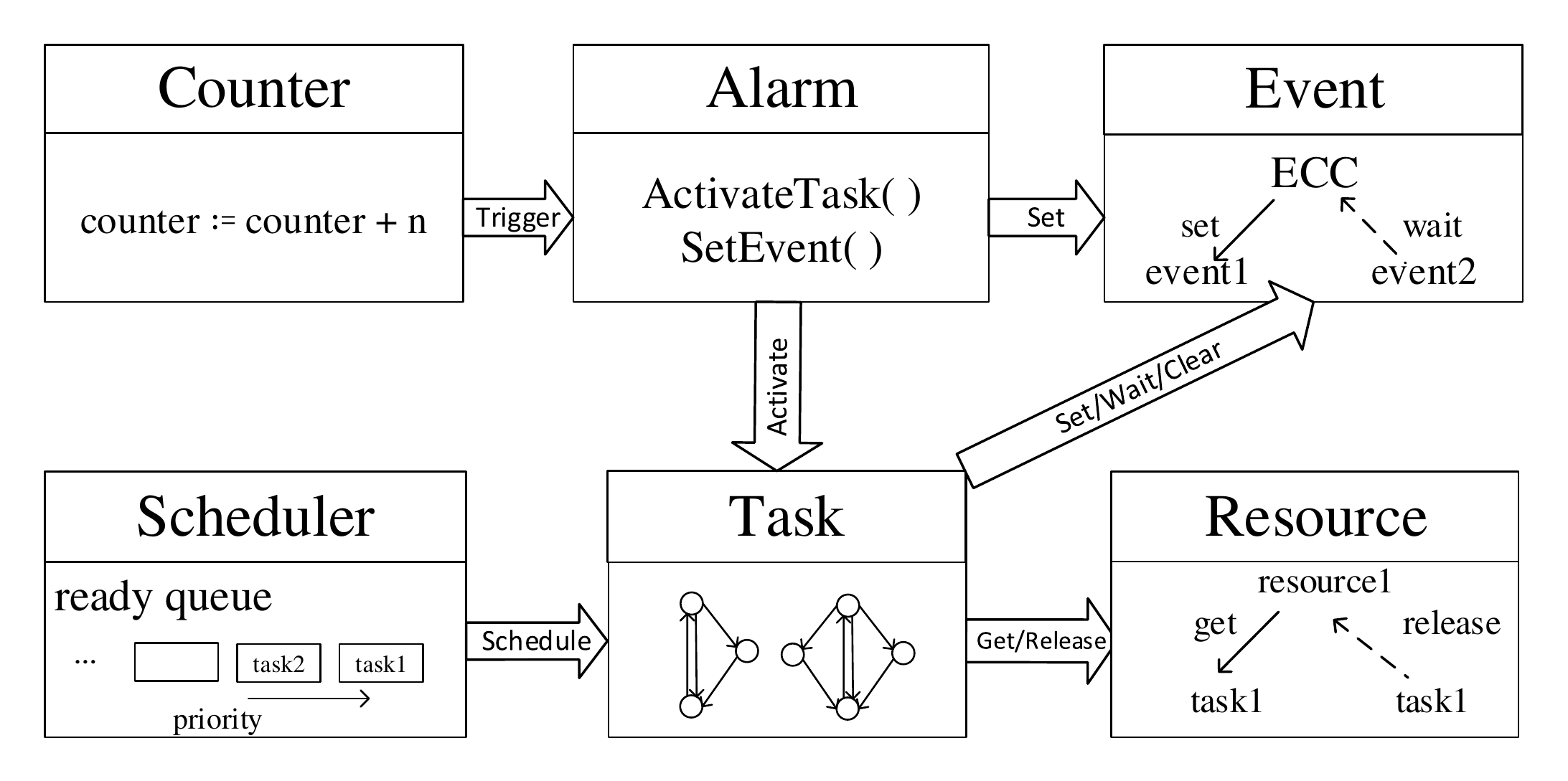}
\caption{Full preemptive scheduling}
\end{flushleft}
\label{fig 1}
\end{figure}

An OSEK/VDX RTOS is a multi-tasking system. This section gives some necessary knowledge of the OSEK/VDX standard for a better understanding of our work. Key conceptions in the standard include task, resource, event, alarms, etc. Task is a basic conception in the standard, which is used to implement complex control software. The execution of tasks is managed by a scheduler on the basis of priority. Tasks can synchronise by the means of events and resources. 
The real time features of OSEK/VDX RTOSs are implemented by alarms instead of timers, which are usually used in embedded operating systems.  Figure \ref{fig 1} shows the relations of these conceptions. 

%
%It can be seen in Figure 1 that the state transition of many modules may activate the scheduler. For instance, based on the priority ceiling protocol, getting or releasing a resource may change the priority of a task, which can activate a scheduler. Another condition to perform scheduling is that an alarm expires and the system calls its service to set an event of activate a task. Alarm mechanism is an important part to achieve time management in the OSEK/VDX-based RTOS. The following part of this section will introduce the alarm mechanism and scheduling policy in detail.

\subsection{Task Scheduling}

\par Scheduler plays an important role in OSEK/VDX RTOSs because it is used to control the execution of tasks. The scheduler decides the next task to run on the basis of its priority with a certain scheduling policy. It is assumed to be fixed-priority \cite{OSEKreview} in that each task should be assigned statically a priority. 
\par Controlled by the scheduler, a task changes between different states. There are two types of tasks, i.e. basic task and extended task, specified in the OSEK/VDX standard. A basic task has three states, i.e., {\it{ready}}, {\it{running}} and {\it{suspended}}. A suspended task is activated to be in {\it{ready}} state, while a running task also can be preempted to be ready. Once a ready task is able to be scheduled, the scheduler allocates it a processor and the task starts running. After finishing, a running task goes into {\it{suspended}} state.
 The extended task has a special state called {\it{waiting}} to indicate the task is waiting for some events. We omitted the state transitions of extended tasks here because it is not the emphasis of the paper. Interested readers are referred to \cite{OSEK} for the details.
\par The scheduler works with a certain scheduling policy. In the standard, there are three scheduling policies specified, i.e., full preemptive scheduling, non preemptive scheduling and mixed preemptive scheduling.

\par By full preemptive scheduling, a running task needs to be rescheduled when a task with a higher priority enters into the ready queue.
We summarise some occasions when a higher-priority task preempts the running task.

\begin{itemize}
\item When a higher-priority task is activated by system services, such as \term{ActivateTask} and \term{ChainTask}.
\item When an extended task with a higher priority is waiting for some event and the event is set by the system service\term{SetEvent}.
\item When the running task releases its resource by the system service \term{ReleaseResource}, and after releasing the resource its current priority becomes lower than some ready task \footnote{The priority of task changes dynamically because the OSEK/VDX RTOS  adopts priority ceiling protocol.}.
\end{itemize}

\par For non preemptive scheduling, rescheduling will not happen unless the state of the running task is changed or a scheduler is called explicitly. Full preemptive scheduling and non preemptive scheduling can coexist in a system, resulting in a mixed preemptive scheduling policy.

\subsection{Alarm}

\par Time management of the OSEK/VDX RTOS is achieved through alarm mechanism. The alarm mechanism is implemented by alarms and counters. A counter, measured by ``ticks", is used to record the number of recurring sources sampled by sensors from the environment.  An alarm is attached to a counter. When the counter reaches a predefined value, the alarm expires and triggers a configured operation. 
In this way, the real time features can be implemented by alarms, such as periodic task or interrupt at regular intervals. 

\par The counters and alarms are configured statically by a system. A counter has three attributes. 
\begin{itemize}
\item MAXALLOWEDVALUE: to specify the maximum value a counter could be. 
\item TICKSPERBASE: to define the count of ``ticks" required by a counter unit.
\item MINICYCLE: to represent the minimum time of ticks for a cyclic alarm.
\end{itemize}
An alarm also has three attributes:
\begin{itemize}
\item COUNTER: to specify the counter to which the alarm is attached.
\item ACTION: to define the type of notification, such as activating a task, setting an event or calling an alarm-callback routine.
\item AUTOSTART: to represent if the alarm is started automatically or not.
\end{itemize}

\begin{figure}[htbp]
\begin{lstlisting}[numbers=left, numberstyle=\scriptsize,xleftmargin=2em]
COUNTER c {
MAXALLOWEDVALUE = 65535;
TICKPERBASE=2;
MINICYCLE=3;
};
ALARM a {
  COUNTER=c;
  ACTION=ACTIVATETASK{
  TASK= t;
  };
  AUTOSTART=false;
};
\end{lstlisting}
\caption{An example of counter and alarm}
\label{fig 2}
\end{figure}
\par 

\par Figure \ref{fig 2} shows an example of counter and alarm. If counter \term{c} reaches its maximum allowed value 65535, it returns to 0 next time. The increment of the counter value is triggered every two ``ticks". If counter \term{c} is attached to a cyclic alarm, the cycle time of the alarm must not be less than 3 ``ticks". In the example, the counter is configured to alarm \term{a}. When the alarm expires, it will activate some predefined task \term{t}. The alarm can not be started automatically and must be set by system services to get started.

\par Two system services can be used to set alarms, i.e., \term{SetRelAlarm (ID, increment, cycle)} and \term{SetAbsAlarm(ID, start, cycle)}. The parameters \term{ID} and \term{cycle} stand for the ID of alarm to be set and its cycle time respectively. If the value of \term{cycle} is  non-zero, the alarm \term{ID} is cyclical. Otherwise, this alarm will expire only once after having been set. In case of calling \term{SetRelAlarm}, the alarm time is the sum of current time and the value of the \term{increment}. 
The parameter \term{start} represents the absolute alarm time without considering current time in \term{SetAbsAlarm}. The system service \term{CancelAlarm(ID)} is used to cancel an alarm according to the standard.

\section{Ambiguities in the OSEK/VDX Standard}
\par The OSEK/VDX standard aims at providing precise statements to represent a uniform environment for efficient implementation of automobile software. However, we found some ambiguous definitions in the process of formalising them. 
In this section, we report and discuss these ambiguities before giving them formal definitions.

\subsection{Ambiguity in Multi-activation of Basic Tasks} According to the standard, there can be multiple requesting of activating a basic task.
The system is able to accept an activation request for a task in the ready or running state in case of the maximum number of activation requests having not been reached.  The maximum number of activation requests is assigned statically.

\begin{figure}[htbp]
\begin{flushright}
\includegraphics[width=0.48\textwidth]{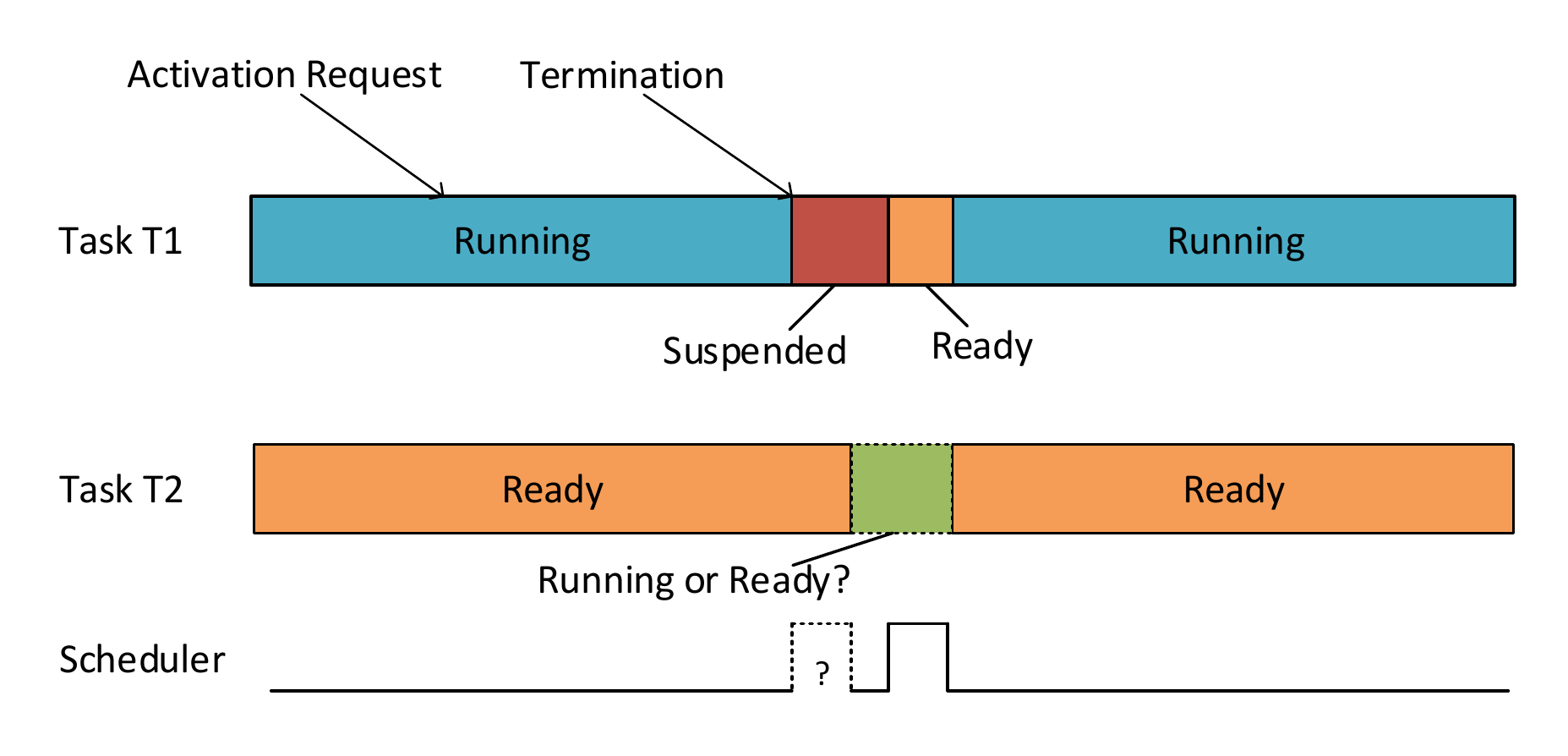}
\caption{ Full preemptive scheduling}
\label{fig 3}
\end{flushright}
\end{figure}
The standard specifies that ``If the task is not suspended, the activation request will only be recorded and performed later" \cite{OSEK}. However, it does not define the exact time when activation request should be handled. Here comes a problem. When the task has been terminated, should the system activate it again first or perform the rescheduling first?

\par Figure \ref{fig 3} illustrates an example of multiple activation request. Both task \term{T1} and \term{T2} are full preemptible and \term{T1} has a higher priority. Task \term{T1} makes an activation request to itself in its {\it{running}} state. It goes into {\it{suspended}} state after being terminated. If the activation request is performed at this time, 
task \term{T1} is transferred to ready state, which triggers rescheduling. Task \term{T1} starts to run due to its higher priority.  If rescheduling is performed before the activation request, task \term{T2} is rescheduled to {\it{running}} state. 
These two cases may lead to different system behaviours. 

\subsection{Ambiguities in System Services about Alarm}
\par We find ambiguities of two system services about alarm in the OSEK/VDX standard, i.e., \term{CancelAlarm} and \term{SetRelAlarm}.
\par  The system service \term{CancelAlarm} is called to cancel an alarm. However, no more information can be found about it in the standard. There can be three understandings of cancelling an alarm leading to different results. 
\begin{itemize}
\item To delete the relation between alarm and its corresponding counter.
\item To set the value of alarm time and cycle time to an unreachable number, such as a number larger than the MAXALLOWEDVALUE.
\item To delete the alarm from the list of working alarms according to the explanation in \cite{program-in-OSEK}.
\end{itemize}

In the first case, the relation can not be set again because the standard does not provide any corresponding system services. In the second case, the alarm needs to be checked every time when the value of the counter is changed although it never expires. This may impact on the execution efficiency. Compared with the first two cases, the third one is more reasonable. Once \term{CancelAlarm} is called, the corresponding alarm becomes invalid. But it can get valid again after being set by other system services.

\par The system service \term{SetRelAlarm} is not clearly defined. According to the standard, its second parameter \term{increment}  stands for the value of relative alarm time. Namely, the alarm time is the sum of current counter value and the \term{increment} value. However, it is specified in the standard that ``The behaviour of \term{increment} equal to 0 is up to the implementation". There are two possible interpretations, i.e., the alarm expires immediately or the first alarm is simply ignored. The second one may cause problems when the third parameter of \term{SetRelAlarm} is 0. Details can be found in Section 4.

\subsection{ambiguous in scheduling policy}

\par The OSEK/VDX standard definitively specifies some conditions of triggering rescheduling \cite{OSEK}. When anyone of the conditions occurs, rescheduling must be activated at once. However, it is possible that some alarm expires after occurrence of the condition but before the activation of rescheduling. It is unclear whether the scheduler should be activated first or not. This may lead to different scheduling results.  
\begin{figure}
\begin{center}
\begin{lstlisting}[numbers=left, numberstyle=\scriptsize,multicols=2,xleftmargin=2em]
Alarm M{
ACTIVATION=
  ACTIVATETASK;{
TASK=C;}
...
};
TASK A{
PRIORITY=1;
AUTOSTART=true;
SCHEDULE=FULL;
...
};
TASK B{
PRIORITY=2;
AUTOSTART=false;
SCHEDULE=FULL;
...
};
TASK C{
PRIORITY=3;
AUTOSTART=false;
SCHEDULE=FULL;
...
};

TASK A{
...
SetRelAlarm(M,1,0);
ActivateTask(B);
TerminateTask();
}
\end{lstlisting}
\caption{An example of application}
\label{fig 4}
\end{center}
\end{figure}
\par Considering a concrete example as shown in Figure  \ref{fig 4}, there are three basic tasks, \term{A}, \term{B}, \term{C} and task \term{C} has the highest priority while the priority of task \term{A} is the lowest. We assume that task \term{A} is currently running and  has just activated task \term{B} (line 28). At this time, the alarm \term{M} expires. If scheduling is handled first, task \term{B} is rescheduled to running state. Then task \term{C} will be activated. Because it has a higher priority than \term{B}, task \term{C} will preempt \term{B} and start to run. If alarm is handled first, rescheduling will be performed after task \term{C} having been activated. Task \term{B} can not run before \term{C} terminates.

\section{ Semantics of the OSEK/VDX Standard}
\par The formal semantics of the standard using  $\mathbb{K}$  framework is presented in this section. The ambiguous definitions of the standard are defined formally after analysing the source code of some certified OSEK/VDX RTOSs developed by iSOFT Infrastructure Software Co., Ltd. \cite{isoft}.
Before presenting the formal semantics, we introduce  $\mathbb{K}$  framework briefly and the major challenges faced when formalising the standard.

\subsection{$\mathbb{K}$  Framework and its underlying Rewriting Logic}

\par $\mathbb{K}$ is a semantic definitional framework based on rewriting logic. Rewriting logic \cite{rewrite} is a computational model for concurrency, distributed algorithms, programming languages, software and hardware systems. It can also be regarded as a logic for executable specification and analysis.  Many languages and systems based on rewriting logic are designed and implemented, such as CafeOBJ \cite{cafeobj}, ELAN \cite{ELAN}, Maude \cite{dmaude} and $\mathbb{K}$, among which,  $\mathbb{K}$  is one of the state-of-the-art frameworks with features of simplicity, executability, analysability, and scalability \cite{Kprimer}.

 $\mathbb{K}$ framework is used to analyse the program languages based on their formal semantics. The syntax of language is defined using BNF ( Backus-Naur Form) in $\mathbb{K}$. And the semantic definitions are like operational semantics, which denote the transition rules of the system. $\mathbb{K}$ framework integrates many capabilities of program analysis. Given a syntax and a semantics of a language, $\mathbb{K}$ generates a parser, an interpreter, a model checker through its Maude backend and a deductive theorem prover by translating its semantics into Coq definitions \cite{coq}. The parser and interpreter allow a program to execute based on its semantics. Furthermore, the $\mathbb{K}$ can analyse the program more comprehensively than compilers. For example, the model checking capability helps the language designer to cover all the non-deterministic behaviors of certain language through the state-space exploration. The main advantages of $\mathbb{K}$ are:
\begin{itemize}
\item $\mathbb{K}$ supports modular definition of formal semantics and user-defined data types.
\item The formal semantics defined in $\mathbb{K}$ is executable, which provides not only a way to formally analyse programs but a way to test the correctness of defined semantics.
\item $\mathbb{K}$ framework has many backend formal tools for interpreting programs, model checking and theorem proving.
\end{itemize}

\par $\mathbb{K}$ is built upon three main components, which are configuration, computational rules and structural rules. The configuration is a nested cell to present the static state of a program being executed. The basic ingredient of a $\mathbb{K}$ definition is cell, which is labeled and stands for a piece of information of a state. The $\mathbb{K}$ rules are executed by changing the information of states stored in cells. Both computational rules and structural rules are $\mathbb{K}$ rules. The difference is that computational rules are counted as computational steps while structural rules are to rearrange the structure of configuration. These concepts will be illustrated in Section 4.3, where we show how the OSEK/VDX standard is formalised in $\mathbb{K}$.

\subsection{Challenges of Formalising the OSEK/VDX Standard}
\par Many formidable challenges have been encountered when formalising the OSEK/VDX standard. The key challenges in this work are about the formalisation of real time features and interrupt.

The first challenge is the formalisation of real time features. In an  automobile operating system time management is an essential part, which has a big influence on task scheduling. It is a key problem to depict real time features for the formal analysis of system behaviors. Time is counted by a physical clock in a vehicle system, which is hard to measure accurately in a laboratory environment. Moreover, $\mathbb{K}$ does not provide any approach to the formalisation of time, and as far as we survey there is no existing work on it.

\par The challenge of depicting time is  to choose an appropriate granularity. In the standard, the increment of a counter may be triggered by a physical clock. The clock time elapses along with the execution of code. Generally, in most of the automobile applications the execution time of an assembly instruction is in microsecond level, while the alarm time is in millisecond. Principally, alarm expiration may occur at any time when an assembly instruction is executed. However, a statement is formalised as an atomic operation in $\mathbb{K}$. Thus it is challenging work to formalise alarm expiration in instruction level. Moreover, we are not concerned with the implementation of system services, but instead just give there formal semantics according to their informal descriptions in the standard. Namely, each system service is considered as an atomic operation. The execution time of a system service, which depends on the amount of its code, the resource occupancy and some other unknown factors, is difficult to evaluate precisely. Meanwhile, we also have to consider the execution time of the statements except system services in tasks, which is another difficulty.

\par Another challenge is the case of interrupt. 
 In the standard, it is not clearly defined whether the execution of system services can be interrupted or not. We find that some system services allow to be interrupted while the other APIs disable the interrupts in its execution in the implementation of a certified RTOS. It means that interrupt can occur during the execution of system service. Therefore, if the system service is considered as an atomic operation in formalisation, some interrupt may be missed. Otherwise, we have to formalise the concrete the implementation of system services. However, how to implement the system service is out of the scope of the standard. Another problem with interrupt is that an interrupt service routine (ISR) may call system services. However, the system services called by ISR are not allowed to be interrupted because the standard specifies that the interrupt can not be nested. These system services should be treated particularly.

\begin{figure*}[htbp]
\begin{center}
$\left\langle\begin{matrix}\big\langle \langle~ List\rangle_{readyTasks} \langle Signal \rangle_{signal} \langle K \rangle_{runningTask}  \langle List \rangle_{alarmList} ...\big\rangle_{global}\\
\big \langle \langle PGM \rangle_{k} \langle TaskState \rangle_{state} \langle N \rangle_{tPriority} \langle N \rangle_{activationTime} \langle N \rangle_{activatedTime}... \big \rangle_{task^*}\\
\big \langle \langle Id \rangle_{aid}  \langle Id \rangle_{aSyscounter} \langle Bool \rangle_{autostart} \langle N \rangle_{alarmTime} \langle N \rangle_{cycleTime}  \langle Bool \rangle_{cyclicity} \langle K \rangle_{action} \langle Id \rangle_{actedId} \langle K \rangle_{actedType}\big\rangle_{alarm}\\
\big \langle \langle Id \rangle_{cid} \langle N \rangle_{maxValue} \langle N \rangle_{cValue} \langle List \rangle_{alarmList}\big\rangle_{sysCounter}......
\end{matrix}\right\rangle_{OSEK}$
\caption{Overview of the Semantic configuration}
\label{fig 5}
\end{center}
\end {figure*}

\subsection{$\mathbb{K}$ Semantics of the OSEK/VDX Standard}
\par The basic idea of formalisation in $\mathbb{K}$ is to represent the system states as configurations and transitions as rewrite rules. 
In an OSEK/VDX RTOS, a system state consists of the information of tasks, alarms, the scheduler and so on. In this work, we mainly consider the transitions of states that are caused through the executions of system services. These transitions are formalised as rewrite rules by defining the semantics of their corresponding system services.

\par In our formalisation, we consider the execution of a system service as atomic due to the fact that the standard is not concerned with the implementation. As discussed in Section 4.2, we need to give a reasonable execution time to each system service. We regard a millisecond as a unit time and consider the execution time of a system service is one unit time in our formal semantics. That is because in general a system service consists of hundreds of assembly instructions and executing an instruction spends several microseconds. Also, in most automobile applications the alarm time are configured in milliseconds. Treating system services as atomic causes the problems mentioned in the second challenge. So far, we have not taken the interrupt process into account and leave it as a piece of future work.

\par In this section, we show the formal definitions of $\mathbb{K}$ configuration of system state, $\mathbb{K}$ semantics of some typical system services on scheduling and alarm mechanism, and the execution time of tasks. We omit the details of the definition of system services on events and resources because it is not the emphasis of this paper.

 \subsubsection{$\mathbb{K}$ Configuration of system state}

\par A configuration in $\mathbb{K}$ consists of possibly nested cells. The $\mathbb{K}$ configuration defined for the standard is composed of over 50 nested cells, describing  necessary information of tasks, alarms, the scheduler and so on. Figure \ref{fig 5} shows an overview of the configuration. The cell \term{task} is declared with multiplicity *, i.e., zero, one, or more tasks can be configured. The cell \term{k} enclosed in \term{task} is the computation cell to store the program to be executed. The \term{activatedTime} keeps the number of task activation requests, and its maximum number is stored in \term{activationTime}. In the \term{alarm} cell, there is a nested cell \term{action} holding the system service, i.e. ActivateTask or SetEvent, to be called by the alarm. The \term{actedId} keeps the identifier of the task to be activated or the event to be set. In the configuration, we define a cell \term{sysCounter} to represent a counter. This cell includes some nested cells to store the information of the counter, such as counter value in \term{cValue}, the alarms it corresponding to in \term{alarmList} and the MAXALLOWEDVALUE in \term{maxvalue}. The TICKSPERBASE is formalises to have the same value with MAXALLOWEDVALUE. We omit its cell  in our semantics.

\begin{figure}[htbp]

RULE {\color{blue}{ACTIVATETASK}}
\\
$\Big\langle~\big\langle\frac{ActivateTask(T);}{.}...\big\rangle_k ~\langle running\rangle_{state}~... \Big\rangle_{task}\\
\Big\langle \langle T\rangle_{tid} \langle N \rangle_{tPriority} \langle \frac{suspended}{ready}\rangle_{state}...~\Big\rangle_{task}\\
 \langle~\frac{L}{add(T, N, L)}~\big\rangle_{readyTasks} \langle AT\rangle_{alarmTime}\\
 \langle \frac{.Set}{SetItem(alarmed)SetItem(schedule)}\rangle_{signal}\\
\Big\langle\big\langle\frac{RV}{(RV+Int~1)\%Int(MAV+Int~1)}\big\rangle_{cvalue}\langle MAV\rangle_{maxValue}\Big\rangle_{counter}$

when $((RV +Int~1)==Int~AT)$\\

RULE {\color{blue}{CHAINTASK-MUL}}
\\
$\Big\langle T\rangle_{tid}~ \langle\frac{running}{suspended}\rangle_{state}~\big\langle\frac{ChainTask(T);}{.}...\big\rangle_k ~\\
~~\langle N1\rangle_{activationTime}~\big\langle\frac{N2}{N2~+Int~1}\big\rangle_{activatedTime}~...~\Big\rangle_{task}\\
\langle AT\rangle_{alarmTime} \big\langle.Set\big\rangle_{signal}\\
\Big\langle\big\langle\frac{RV}{(RV+Int~1)\%Int(MAV+Int~1)}\big\rangle_{cvalue}\langle MAV\rangle_{maxValue}\Big\rangle_{counter}$

when $N2 <Int$ N1 andBool $((RV +Int~1)<Int~AT)$\\

RULE {\color{blue}{MULTIACTIVATED}}
\\
$\Big\langle T\rangle_{tid}~ \langle N\rangle_{tPriority}~\big\langle\frac{suspended}{ready}\big\rangle_{state}\\
~~~\big\langle\frac{N1}{N1-1}\big\rangle_{activatedTime}~...~\Big\rangle_{task}\\
~~~\big\langle~\frac{L}{add(T, N, L)}~\big\rangle_{readyTasks}\big\langle\frac{.Set}{SetItem(shedule)}\big\rangle_{signal}$

when N1$=/=Int$ 0\\

%RULE ADD1\\
%\\
% $\frac{add(I, ~N, ~.ListTaskPair)}{<I; ~N>~.ListTaskPair}$\\
%\\
%RULE {\color{blue}{ADD}}\\
%~~~ $\frac{add(I, ~N, ~(<I';~N'>~L))}{<I; ~N>~(<I';~N'>~L)}$
%
%when N$>Int~N'$\\
%\\
%RULE ADD3\\
%\\
% $\frac{add(I, ~N, ~(<I';~N'>~L))}{<I'; ~N'>~add(I,~N,~L)}$\\
%\\
%when N$<=Int~N'$

\caption{Main rewrite rules for scheduling}
\label{fig 6}

\end{figure}

\subsubsection{$\mathbb{K}$ semantics of system services on scheduling}\label{subsec:multi}
\par There are several system services that can trigger scheduling as mentioned in Section 2.1. We consider two typical system services \term{ActivateTask} and \term{ChainTask} as examples here.
The two system services have different semantics on full preemptive tasks and non preemptive tasks, which should be formally defined respectively. The semantics on full preemptive tasks is shown in Figure \ref{fig 6}.

\par There are six occasions when executing the system service \term{ActivateTask(T)} leads to different results. 
 The first rule [{\color{blue}{ACTIVATETASK}}] in Figure \ref{fig 6} formalises the semantics of one occasion when the alarm expires.  According to the standard, if \term{ActivateTask(T)} is called, the state of task \term{T} is transferred from {\it{suspended}} to {\it{ready}} as specified by the \term{state} cell of the second \term{task}. In the cell, the horizontal line indicates a reduction.  The cells without horizontal lines are only read but do not change during the reduction. For instance, the cell \term{state} represents the task that is calling \term{ActivateTask(T)} is in {\it{running}} state. 
The system service \term{ActivateTask(T)} is consumed after having been executed. It is represented in cell \term{k} by the reduction from \term{ActivateTask(T)} to ``.", which stands for empty or the unit computation. The ``..." are structural frames, which match the irrelevant portions of this cell. 
Because task \term{T} gets ready, it is added to the corresponding queue of ready tasks, which have the same priority \term{N} as \term{T}. This is specified by the cell \term{readyTasks}, where term \term{L} is reduced to \term{add(T, N, L)}. \term{L} is a variable for the list of queues of ready tasks, and \term{add(T, N, L)} returns a new list by adding \term{I} to the queue in \term{L} with the priority \term{N}.

\par Executing the system service causes one-unit increment to the counter value. The value of counter needs to begin from 0 after it reaches it maximum allowed value, which is formalised by the reduction from \term{RV} to \term{(RV+Int 1)\%Int (MAV+Int 1)}. \term{RV} and \term{(RV+Int 1)\%Int (MAV+Int 1)} are two terms to represent the counter values before and after the system service is executed respectively. \term{MAV} is a variable for the maximum allowed value of a counter,   and \term{+Int}, \term{\%Int} are two built-in operators for addition and modulo operations on integers. 

\par Every time when the value of counter is changed, alarm expiration needs to be checked. In this rule, \term{AT} in cell \term{alarmTime} expresses the time when alarm expires. The expression \term{RV +Int 1==Int AT} following keyword \term{when} represents the condition of alarm expiration. On reaching the expiration time, a signal called \term{alarmed} is released. Meanwhile, the signal \term{schedule} is also released which represents that the scheduler needs to be triggered after the execution of \term{ActivateTask(T)}. The release of these two signals is formalised by cell \term{signal}. The handling of these two signals is to be introduced in Section 4.3.3.

\par The second rule in Figure \ref{fig 6} specifies the semantics of \term{ChainTask (T)} in the case of multiple activation request.  This system service may cause the termination of the running task and the activation of a task according to the OSEK/VDX standard. The rule is defined for the case that the task to be activated is just the task in running state.
Namely, task \term{T} is requested to be multiply activated. Based on the analysis in Section 3.1, there can be two possible methods for handling the multiple activation request. In the second method, i.e., performing the scheduling first, 
it is unclear when the activation should be performed resulting in multiple interpretations on the activation time. In the first method, the activation time is clearly specified. 
Therefore, we adopt the first method, which means that no \term{schedule} signal is released, though the task \term{T} is terminated.

\par The rule has a condition \term{N2<Int N1} meaning that the number of activation request of task should not exceed its maximum value. When the condition is met and the alarm does not expire, the activation time is increased by one as represented in cell \term{activatedTime}.

\par Rule [{\color{blue}{MULTIACTIVATED}}] formalises the handling of the activation information in the case that a multi-activated task \term{T} with priority {N} has transferred to \term{suspended} state.
The state of task \term{T} is transferred to \term{ready} as soon as it is terminated, which triggers the \term{schedule} signal. Meanwhile, the activation time of this task is decreased by one, i.e. the value of \term{N1} decreases to \term{N1-1} in cell  \term{activatedTime}. %After task \term{T} gets ready, it is added to the corresponding queue of ready tasks, which have the same priority \term{N} as \term{T}. This is specified by the cell \term{readyTasks}, where term \term{L} is reduced to \term{add(T, N, L)}. \term{L} is a variable for the list of queues of ready tasks, and \term{add(T, N, L)} returns a new list by adding \term{I} to the queue in \term{L} with the priority \term{N}.

\begin{figure}[htbp]

RULE {\color{blue}{SETRELALARM}}
\\
$\Big\langle~\langle running\rangle_{state}~\big\langle\frac{SetRelAlarm(I, N1, N2);}{.}...~\big\rangle_k...~\Big\rangle_{task}\\
\Big\langle~\langle I\rangle_{aid}~\big\langle\frac{-}{(N1~+Int~RV)~\%Int~(MAV~+Int~1)}\big\rangle_{alarmTime}\\
~~~~\big\langle\frac{-}{N2}\big\rangle_{cycleTime}\big\langle\frac{-}{true}\big\rangle_{cyclicity}\langle C\rangle_{aSysCounter}...\Big\rangle_{alarm}\\
\Big\langle~\langle C\rangle_{cid}~\big\langle\frac{RV}{(RV~+Int~1)~\%Int~(MAV~+Int~1)}\big\rangle_{currentValue}\\
~~~~\langle MAV\rangle_{maxAllowedValue}...\Big\rangle_{sysCounter}\\
~~~~\big\langle\frac{L}{L~ListItem(I)}\big\rangle_{alarmList} \big\langle\frac{.Set}{SetItem(alarmed)}\big\rangle_{signal}$\\
when N2 $=/=Int$ 0 andBool N1$==Int$ 0\\
\\
RULE {\color{blue}{CANCLEALARM}}
\\
$\Big\langle~\langle running\rangle_{state}~\big\langle\frac{CancelAlarm(I);}{.}...~\big\rangle_k...~\Big\rangle_{task}\\
\Big\langle~\langle I\rangle_{aid}~\big\langle\frac{-}{(N1~+Int~RV)~\%Int~(MAV~+Int~1)}\big\rangle_{alarmTime}...\Big\rangle_{alarm}\\
\big\langle\frac{L}{remove(I, L)}\big\rangle_{alarmList} \langle.Set\rangle_{signal}$\\

RULE {\color{blue}{TWO-SIGNALS}}
\\
~~~$\big\langle\frac{SetItem(alarmed)~SetItem(schedule)}{SetItem(schedule)}\big\rangle_{signal}\langle L'\rangle_{alarmList}\\
\Big\langle~\langle T\rangle_{tid}~\big\langle\frac{suspended}{ready}\big\rangle_{state}~\langle N\rangle_{tPriority}...\Big\rangle_{task}\\
~~~\langle\frac{L:ListTaskPair}{add(T, N, L)}\rangle_{readyTask} \langle RV \rangle_{cValue}\\
\Big\langle~\langle I\rangle_{aid}\big\langle\frac{AT}{AT~+Int~CT}\big\rangle_{alarmTime}\langle CT\rangle_{cycleTime}\\
~~~~\langle ACTIVATE\rangle_{action}\langle T\rangle_{actedTaskId}\\
~~~~\langle task\rangle_{actedType} \langle true \rangle_{cyclicity}...\Big\rangle_{alarm}$\\
when (I in L') andBool (AT==Int RV)\\
\\

\caption{Main rewrite rules for alarm mechanism}
\label{fig 7}

\end{figure}
\subsubsection{$\mathbb{K}$ Semantics of Alarm}\label{subsec:alarm}

\par We formally define 12 rules for three system services on alarms and another 4 rules to specify the handling of \term{alarmed} signal. Figure \ref{fig 7}  shows three typical rules of them.

\par The first rule [{\color{blue}{SETRELALARM}}]  formalises one case of the system service \term{SetRelAlarm(I, N1, N2)}. According to the standard, this system service is to start alarm \term{I} and set its relative alarm time as \term{N1} and cycle time as \term{N2}. After the system service, the alarm time should be \term{N1+Int RV}. 
 Besides, if the alarm time is greater than the maximum allowed value of the counter (\term{MAV}), it is becomes the value modulo \term{MAV+Int 1}. Namely, the alarm time is reduced to \term{(N1+Int RV)\%Int (MAV+Int 1)} as represented in cell \term{alarmTime}. 
When \term{N1} is 0, we discussed two possible interpretations of \term{SetRelAlarm(I, N1, N2)} in Section 3.2. 
By the second interpretation, the first alarm is simply ignored. However, if the value of \term{N2} is 0 this alarm can never take effect because \term{N2} being 0 means the alarm is not cyclic and just alarms once. In this sense, the first interpretation is more reasonable. Therefore, we take the first interpretation in our formal semantics, i.e. the alarm \term{AlarmID} expires immediately and \term{alarmed} signal is released, as represented in cell \term{signal} of this rule. After the alarm \term{I} is set it is added to current alarm list \term{L}, which is specified in cell \term{alarmList}.

\par The rule  [{\color{blue}{CANCELALARM}}] specifies the execution of system service \term{CancelAlarm(I)}. We have discussed three possible understandings to cancel an alarm in Section 3.2 and concluded that the third one is more reasonable. Therefore, we adopt the third understanding in our formalisation, that is, the alarm \term{I} is cancelled by moving from the alarm list in cell \term{alarmList}.

%Its \term{alarmTime} is set as \term{(N1 + RV) \%Int (MAV + 1)}, here \term{N1} stands for the relative alarm time of alarm \term{I}, RV is the current time of the counter. In this term, "-" is an anonymous variable, which can match anything. Here this variable means we don't care what the initial value of \term{alarmTime} at all. The reason for modulo \term{MAV + 1} is that the maximum number of the counter is \term{MAV}. When the counter's value reaches \term{MAV}, it is counted from 0 next time. \term{N2} represents the cycle length of this alarm. Assume that the first alarm time is N, then the next alarm time is N+N2, the third alarm time is N+N2+N2 and so on. 

\par  Rule [{\color{blue}{TWO-SIGNALS}}] formalises a case of the handling of signals. When both of the signal  \term{schedule} and \term{alarmed} are triggered at the same time, there are two possible handling ways as discussed in Section 3.3. We first handle the signal \term{alarmed} on the basis that alarm is considered as a kind of interrupt in \cite{OSEKreview} and the priority of interrupts is the highest with respect to the OSEK/VDX standard. In the rule, the condition specifies that the signal is released by alarm \term{I}.
The cell \term{alarm} shows that the action of \term{I} is to activate task \term{T}. The formalisation of this action is same as the one of system service \term{ActivateTask(T)}. We omit detailed explanations here. The cell \term{cyclicity} shows that alarm \term{I} is cyclic. Thus its next alarm time becomes \term{AT+CT} as shown in cell \term{alarmTime}. After the signal \term{alarm} is handled it is removed from the cell \term{signal}.
% In this  rule, the signal \term{schedule} won't be dealt with. Besides, a new task is inserted into the queue of ready task. Thus, the data in the \term{signal} is rewritten into \term{SetItem(schedule)}. Here the \term{schedule} of \term{SetItem(schedule)} is an element of a set. The difference between set and list is that the elements in the set is unordered and different. Thus, 

\subsubsection{Formalisation of the execution time of tasks}

\par As mentioned in the first challenge in Section 4.2, we need to take the execution time of the statements except system services in tasks into account. Because our work mainly focuses on the real time features, for simplicity we do not deal with the formal semantics of these statements but formalise their execution time. System services partition the statements in a task into several blocks. Instead of measuring each statement, we consider a block as a whole to evaluate its execution time. For the reason that in our formalisation the execution time of a system service is considered as one unit time, the  execution time of a block is measured through the value that results from dividing the amount of statements in the block by the average amount of statements in system services. 
\par In general, it is difficult to calculate the amount of statements in a block due to the fact that there may be recursive function calls, loops with uncertain times and so on. However, our survey shows that only simple statements, such as assignments, \textbf{if} statements and fixed-time loops and non-recursive function calls, are used in the industrial implementations of a task. Therefore, we assume that a block only consists of the above four kinds of statements. For \textbf{if} statements, we further assume that each branch has the same execution time and then the execution time of an \textbf{if} statement is equal to the one of its branches. For fixed-time loops and non-recursive function calls, we calculate the amount of statements by loop unrolling and function inlining.

\par In order to specify the execution time of a block, we introduce a special statement in form of \term{TimeInterval=N}  to replace the block, where \term{N} stands for the execution time. The semantics of this statement is to add \term{N} to the counter value. The new value of counter has two possibilities, i.e., it exceeds alarm time or not. If the counter value does not exceed the alarm time, \term{N} is directly added to the counter value as shown in rule  [{\color{blue}{TIMEINTERVAL2}}] of Figure \ref{fig 8}.
To deal with the other case, we need to divide \term{N} into two time intervals such that after the first time interval alarm happens to expire. Namely the first time interval is equal to the alarm time minus counter value. After the first time interval, we handle the alarm and then treat the second time interval as an independent one to deal with recursively. The handling of this case is formalised by  rule  [{\color{blue}{TIMEINTERVAL1}}]. In this way, we guarantee that no alarm expiration is missed during the execution of a block, which complies with the fact that alarm may expire during the execution of statements in a task.

\begin{figure}[htbp]
RULE {\color{blue}{TIMEINTERVAL1}}
\\
$\big\langle\frac{TimeInterval=N;}{TimeInterval=((N~+Int~RV)-Int~AT);}...\big\rangle_k
\\
\Big\langle\langle A\rangle_{aid}\langle C\rangle_{aSysCounter}\langle AT\rangle_{alarmTime}...\Big\rangle\\
\Big\langle\langle C\rangle_{cid}\big\langle\frac{RV}{AT~\%Int~(MAV~+Int~1)}\big\rangle_{currentValue}\\
~~~\langle MAV\rangle_{maxAllowedValue}...\Big\rangle_{sysCounter}\\
~~~\langle L\rangle_{alarmList}\big\langle\frac{.Set}{SetItem(alarmed)}\big\rangle_{signal}\\
when~(A~in~L)~andBool~((RV~+Int~N)~>=Int~AT)$\\

RULE {\color{blue}{TIMEINTERVAL2}}
\\
$\big\langle\frac{TimeInterval=N;}{.}...\big\rangle_k
\\
\Big\langle\langle A\rangle_{aid}\langle C\rangle_{aSysCounter}\langle AT\rangle_{alarmTime}...\Big\rangle\\
\Big\langle\langle C\rangle_{cid}\big\langle\frac{RV}{AT~\%Int~(MAV~+Int~1)}\big\rangle_{currentValue}\\
~~~\langle MAV\rangle_{maxAllowedValue}...\Big\rangle_{sysCounter}\\
~~~\langle L\rangle_{alarmList}\big\langle.Set\big\rangle_{signal}\\
when~(A~in~L)~andBool~((RV~+Int~N)~<Int~AT)$\\

\caption{Main rules for the formalisation of execution time}
\label{fig 8}

\end{figure}

\section{Application to Conformance Verification}

\par The $\mathbb{K}$ semantics can be used for system analysis in various ways, such as model checking, state space exploration and test case generation. The work \cite{OSEK-K} shows these applications with the $\mathbb{K}$ semantics of the OSEK/VDX standard. In this section, we propose another application to system analysis with the $\mathbb{K}$ semantics called {\it{conformance verification}} and show a case study on it.

% how our semantics together with builtin $\mathbb{K}$ tools can be used to verify
%
%as an simulater and model checker to explore an application behaviours and verify LTL properties. The OSEK/VDX-based application is configured in OIL (OSEK Implementation Language) \cite{OIL} and implemented in C-like language. The goal of OIL is to provide a standard mechanism to configure the resources, tasks and some other objects in an OSEK/VDX-based operating system.
\subsection{The Basic Idea of Conformance Verification}
\par Automobile operating systems are required to conform to the standard in order to get certified by the OSEK/VDX organisation. The traditional  technique for ensuring the conformance is conformance testing. A conformance testing method named MODISTARC \cite{modistarc} has been proposed for the OSEK/VDX RTOS. The basic idea is to test the result of running an operating system kernel with a given suites of test cases in a given environment. The organisation certifies an operating system if it can pass the testing. This method is practical but has some limitations due to the testing technique. In addition, only system services in operating system kernel are tested in the method while an automobile operating system including its applications as a whole is supposed to conform to the standard.

%However, the MODISTARC also exists the problems mentioned in Section 1due to its informal methodology. Besides, this method can just test the conformance of the operating system kernel. As a part the automobile operating system, the applications also needs to comply with its standards, i.e. OSEK/VDX and OIL (OSEK Implementation Language) standard. The OIL provide a standard mechanism to configure the resources, tasks and some other objects in an OSEK/VDX operating system.

\begin{figure}[htbp]
\begin{center}
\includegraphics[width=0.5\textwidth]{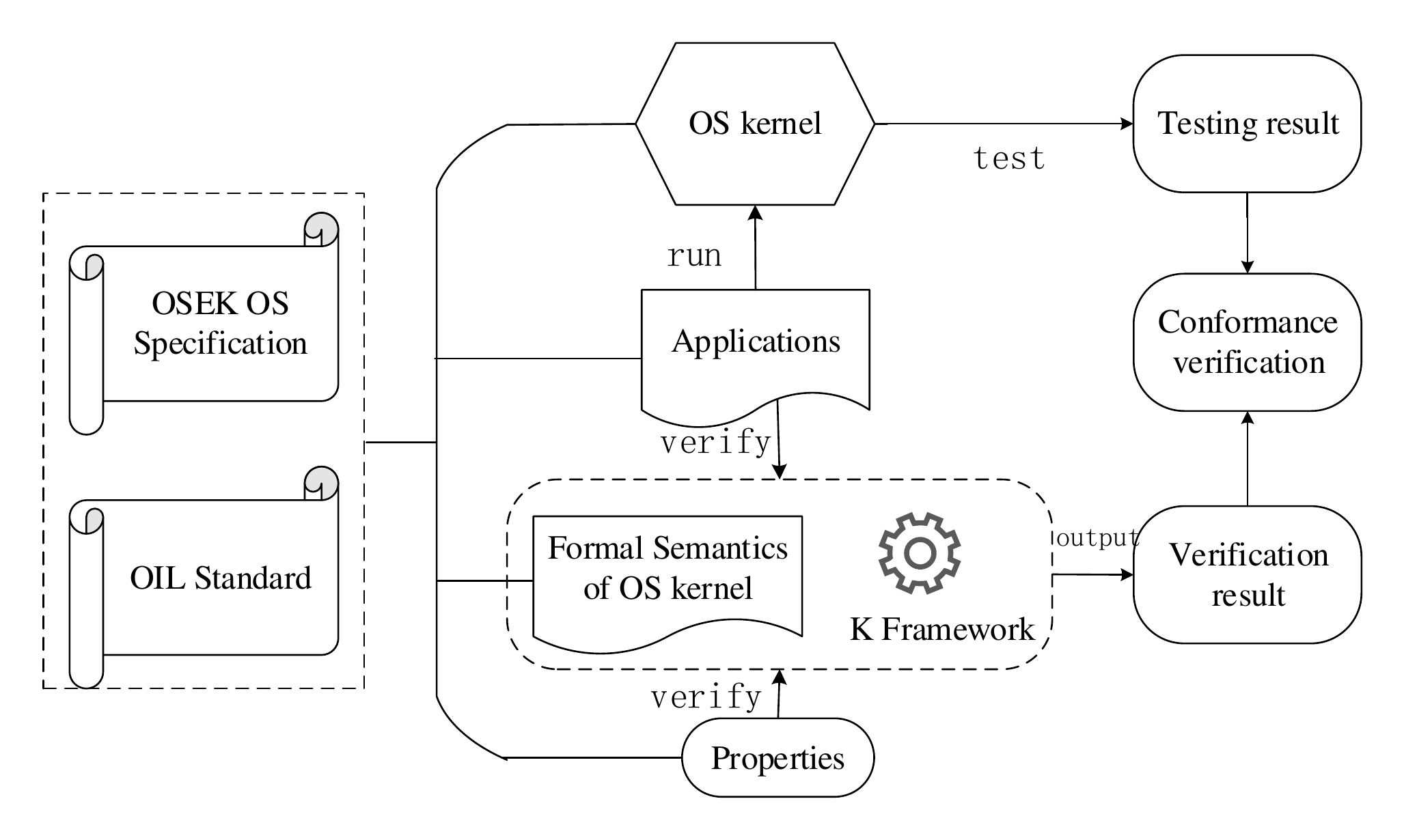}
\caption{\label{fig 9}The architecture of conformance verification approach}
\end{center}

\end{figure}

\par As a complementary approach, conformance verification is used to check the conformance of not only the operating system kernel but the applications. Figure 9 shows the architecture of the approach to conformance verification. 
\par Firstly, we define the $\mathbb{K}$ semantics of the OSEK/VDX and OIL  standards \cite{OIL}. The OIL is abbreviated for OSEK Implementation Language, used to configure resources, tasks and some other objects in  OSEK/VDX applications. Then we extract some properties from the standards and formalise them as LTL formulae \cite{MC}. These properties are supposed to be satisfied by any automobile operating system. Given a concrete implementation, we identify from the requirement the properties that should be satisfied by the applications. Next we verify the two kinds of properties by model checking the applications with the $\mathbb{K}$ semantics of the OSEK/VDX and OIL standards. And we test these properties by executing the applications on the operating system kernel. Finally, we conclude conformity or inconformity of operating system kernel and applications on the basis of the verification and testing result.

\par Under the assumption that the formal semantics is correct, we obtain the relations between verification \& testing result and conformance verification result as shown in Table \ref{tab 2}. If the verification result of a property is consistent (or inconsistent) with the testing result, we can conclude the conformance (or unconformance) between the kernel and the standard with respect to the property. The conformance of applications can be directly concluded by the verification result.

\begin{table}
\begin{center}

\begin{tabular}{|p{1.5cm}|p{1.5cm}|p{1.5cm}|p{1.5cm}|}
\hline 
\multicolumn{2}{|c|}{Verification \& Testing Result} & \multicolumn{2}{|c|}{Conformance Verification Result} \\
\hline
\multicolumn{1}{|c|}{verification} & \multicolumn{1}{|c|}{testing} & {conformance   $~~~~$of kernel} & {$~~$conformance   $~~~~~~~~$of app} \\
\hline
 \multicolumn{1}{|c|}{pass} &  \multicolumn{1}{|c|}{pass} &  \multicolumn{1}{|c|}{$\surd$} &  \multicolumn{1}{|c|}{$\surd$} \\
\hline
 \multicolumn{1}{|c|}{pass} &  \multicolumn{1}{|c|}{fail} &  \multicolumn{1}{|c|}{$\times$} &  \multicolumn{1}{|c|}{$\surd$}  \\
\hline
 \multicolumn{1}{|c|}{fail} &  \multicolumn{1}{|c|}{pass} &   \multicolumn{1}{|c|}{$\times$} &  \multicolumn{1}{|c|}{$\times$} \\
\hline
 \multicolumn{1}{|c|}{fail} &  \multicolumn{1}{|c|}{fail} &   \multicolumn{1}{|c|}{$\surd$} &  \multicolumn{1}{|c|}{$\times$} \\
\hline
\end{tabular}

\begin{tablenotes}
        \footnotesize
        \item $\times$: inconformity \quad $\surd$: conformity  
      \end{tablenotes}
%    \end{threeparttable}   
    \end{center}
   \setlength{\abovecaptionskip}{5pt}
\setlength{\belowcaptionskip}{15pt}
\caption{ \label{tab 2} Relations between verification \& testing result and conformance verification result}

  \end{table}

\par The approach is feasible to automobile systems though the conformance verification result relies on the properties of applications. That is because the automobile system is a closed system in that no applications are allowed to install after it is deployed. It suffices to verify the conformance based on properties of the existing applications in an automobile system.

%\par The key process of our approach to conformance verification is shown in Figure \ref{fig 9}. Firstly, we analyse the OSEK/VDX specification and OIL standard to define the formal semantics of tasks, events, resources, alarms and counters. The formal semantics can be executed with $\mathbb{K}$ Framework as an interpreter or model checker to test or verify the applications. An automobile application is made up of two parts, which are configuration file and implementation file. The configuration file predefines the static attributes of components, such as priority of tasks, in OSEK/VDX-based application. Tasks in the application are implemented in C-like imperative programming language. The testing and verification results can provide a reliable feedback to the applications and specifications.
%The applications or specifications 
%
%The execution of an application is controlled by the implementation file. This file calls some functions and some system services to achieve the goal of the application. The application can be tested by our framework. If the execution result is not the expected one, there may be some bugs in the application or some errors in the specifications. Besides, the application with some given properties can be verified in our framework. The 

\subsection{Case Study}

\par We conduct a case study on an industrial RTOS and show how the conformance verification can be achieved by adopting an application called EMS, which is built in the system.

\subsubsection{Overview of EMS application}
\par EMS is one of the most important applications in the automotive operating system, which requires strict time constraints. It communicates with the operating system kernel by CAN (controller area network) bus. Because we are concerned with the real time features of the automobile system, we ignore the bus mechanism and assume a direct communication between EMS and the kernel.

\par EMS consists of two parts, i.e., configuration and implementation. The configuration specifies the statical attributes of alarms, tasks, events and so on. 
In EMS there are five preemptible tasks and they can not be multiply activated. The sketch of their implementation is shown in Figure \ref{fig 10}. Task \term{SystemInit} is set as an autostart task in the configuration file \footnote{We omit the configuration file of the application due to the space limitation.}. It sets three alarms which are used to activate three tasks periodically. Task \term{AL\_Task\_10ms} is responsible for sampling A/D data and is activated with the period of 10 ms. Task \term{AL\_EMS\_Task\_100ms}  is used to get the rotational speed of the engine and is activated per 100 ms. Task \term{AL\_EMS\_Task\_10ms} updates the engine state and triggers the combustion process in four cylinders every 10 ms.  All the three tasks must be finished in a certain time. Otherwise they may be multiply activated, which violates the configuration and hence results in unexpected behaviors.
\par This work mainly focuses on the real time features of the system that rely on the execution time of not only system services but other statements in tasks. As mentioned in Section 4.3.4, we abstract the statements except system services as a special statement in form of \term{TimeInterval=N} to express their execution time. For instance, there are about 360 statements in the block of task \term{EMS\_Task\_10ms} after loop unrolling and function inlining, and the average amount of statements per system service is 110. Therefore, the execution time is 360/110, amount to 3 units time. We represent it  as \term{TimeInterval=3} (line 16 in Figure \ref{fig 10}).

%The task \term{SystemInt} has the highest priority in this application if it is conformant  to the OSEK/VDX standard.

\begin{figure}[htbp]
\begin{lstlisting}[numbers=left, numberstyle=\scriptsize,xleftmargin=2em]
int a; 
/*PRIORITY=3*/ 
TASK EMS_Adap_Task_10ms{
 while(true){
  WaitEvent(Adap_Event);
  ClearEvent(Adap_Event); 
  }
  TerminateTask();
};
/*PRIORITY=2*/
TASK EMS_Task_100ms{
  TerminateTask(); 
};
/*PRIORITY=1*/
TASK EMS_Task_10ms{
  TimeInterval = 3;
  TerminateTask(); 
};
/*PRIORITY=0*/
TASK Task_10ms{	
  SetEvent(EMS_Adap_Task_10ms, Adap_Event);  
  TerminateTask(); 
};
/*PRIORITY=4
  AUTOSTART=1*/
TASK SystemInit{
  SetRelAlarm(AL_Task_10ms,6 , 10);
  /*ActivateTask(Task_10ms)*/
  SetRelAlarm(AL_EMS_Task_100ms, 7, 100);
  /*ActivateTask(EMS_Task_100ms)*/
  SetRelAlarm(AL_EMS_Task_10ms, 8, 10);
  /*ActivateTask(EMS_Task_100ms)*/
  ActivateTask(EMS_Adap_Task_10ms);	
  TerminateTask();
};
\end{lstlisting}
\caption{\label{fig 10}Sketch of EMS implementation part}

\end{figure}

\subsubsection{Conformance Verification}

\par We verify the conformance between the industrial RTOS and the OSEK/VDX standard with respect to EMS. As depicted in Figure \ref{fig 9}, we identify six properties from the standard and the application, and then verify them with the $\mathbb{K}$ semantics of the standard. Two of them are verified invalid but they pass the testing on the RTOS. By the result we find two inconformities in the implementation.

\par The six properties used for conformance verification are listed below. The first three are extracted from the standard and the other three are extracted from the requirement of EMS. 
\begin{itemize}
\item Deadlock Freedom (DF): The system never reach a state from which no task can run.
\item Mutual Exclusion (EX): At any moment, there is at most one task in running state.
\item Priority Inversion Freedom (PIF): If a task becomes ready and its priority is higher than the one of the running task, it must preempt the running task next step. 
\item Starvation Freedom (SF): The task that calls system service \term{WaitEvent(E)} must have been or will be set event \term{E} .
\item Periodic Execution (PE): A periodic task should be executed once at its specified interval. 
\item Multiple-Activation Freedom (MAF): The task that is configured to be one-shot activation must not be activated in its running or ready state.

\end{itemize}

\par We formalise the above six properties and verify them using the built-in toolkit of K. For instance, deadlock freedom can be verified by exploring the state space of EMS. The command used for the verification is as follows:

\begin{lstlisting}[language={[ANSI]C},keywordstyle=\color{blue!70},commentstyle=\color{red!50!green!50!blue!50},frame=shadowbox, rulesepcolor=\color{red!20!green!20!blue!20}] 
krun EMS.osek --search-final
\end{lstlisting}

\term{krun} is a $\mathbb{K}$ command, used to execute a specified program, e.g. \term{EMS.osek}, with the $\mathbb{K}$ semantics of the language. The option \term{--search-final} of \term{krun} indicates to exploring all the possible executions and returning final states. There are two possibilities with final states, i.e. occurrence of deadlock or reaching the specified bound. However, in the task \term{EMS\_Adapt\_Task\_10ms} there is an infinite loop (line 4 in Figure \ref{fig 10}), which leads to non-termination of execution. We deal with it by setting a bound. 
 
\par As another example, we consider the verification of the starvation freedom using LTL model checking in $\mathbb{K}$. The automobile operating system does not allow a task always being in {\it{waiting}} state. In EMS, when task \term{EMS\_Adap\_Task\_10ms} waits for some events, at least one of the events it is waiting for must be set to \term{EMS\_Adap\_Task\_10ms} sometime. This property can be described as the following LTL formula:\\

%There are three periodic tasks to be verified. We take task \term{Task\_10ms} for instance. According to the requirement, the task is supposed to be activated every 10 unit time after the sixth. The property for task \term{Task\_10ms} 
\begin{description}
\item $\Box ( wait(Adap\_Event, EMS\_Adap\_Task\_10ms) \rightarrow \\ \Diamond set(Adap\_Event, EMS\_Adap\_Task\_10ms))$
\end{description}

Here \term{ wait(Adap\_Event, EMS\_Adap\_Task\_10ms)} is a Boolean function with parameters \term{Adap\_Event} and \term{EMS\_Adap\_Task\_10ms}, which returns true when task \term{EMS\_Adap\_Task\_10ms} calls the system service \term{WaitEvent(Adap\_Event)}, and false otherwise. The function \term{set(Adap\_Event, EMS\_Adap\_Task\_10ms)} returns true when the event \term{Adap\_Event} is set to \term{EMS\_Adap\_Task\_10ms}.

%Here \term{counterValue(6, 10)} is a Boolean function with parameters 6 and 10, which returns true when the counter value modulo 10 equals 6, and false otherwise. The function \term{running(Task\_10ms)} (or \term{suspended(Task\_10ms)}) returns true when the task is in {\it{running}} state (or in {\it{suspended}} state). Figure \ref{fig 11} illustrates graphically the LTL formula. \\

The starvation freedom property can be verified through LTL model checking in $\mathbb{K}$ \cite{MC} with the following command:
\begin{lstlisting}[language={[ANSI]C},keywordstyle=\color{blue!70},commentstyle=\color{red!50!green!50!blue!50},frame=shadowbox, rulesepcolor=\color{red!20!green!20!blue!20}] 
krun EMS.osek --ltlmc ltlformula
\end{lstlisting}
The option \term{--ltlmc} indicates that a specified program, e.g. \term{EMS.osek}, is model checked with the LTL formula \term{ltlformula}. A counterexample is returned when \term{ltlformula} fails to be verified.

\subsubsection{Verification Result}

\par Table \ref{tab 1} shows the verification and testing results of the six properties. The results of the properties DE and MAF are inconsistent while those of other four properties are consistent. By Table \ref{tab 2} we can conclude that the operating system kernel and EMS are unconformant to the standard with respect to the properties DE and MAF. 
\par We obtain two counterexamples when model checking the two properties. The counterexample of DE deadlock occurs when the value of counter is 16. In this time the alarm \term{AL\_Task\_10ms} expires and is expected to activate task \term{Task\_10ms}. However, this task is in ready state and can not be activated because it is configured as a one-shot activation task. Hence the error occurs. 
\par By analysing the implementation of EMS, We notice that the task \term{Task\_10ms} has the lowest priority but the shortest period. It violates the general priority configuration criterion, that is, a shorter period task has a higher priority. After adjusting the priorities of the tasks in EMS under the criterion, the two properties are verified to be valid. Because EMS with original priorities passes the test on the kernel, there may be potential bugs in the kernel. A possible approach to detecting these bugs is to compose test cases according to the returned counterexamples and use them to test the kernel.

%\par In the verification, the two properties DE and MAF fail and the other four are verified valid. However, the industrial RTOS is tested to satisfy the six properties by running EMS on the kernel. Table \ref{Tab 1} shows the verification and testing results, which obviously is inconsistent. 
%
%\par There are two possible causes to the inconsistency between the verification and testing results. One is the RTOS indeed satisfies with the two properties DE and MAF. The other is that the RTOS actually violates them, but no corresponding bugs are detected due to the limitation of testing. 

\begin{table}

\begin{center}
\begin{threeparttable}
\begin{tabular}{|>{\centering}p{1.2cm}|c|>{\centering}p{2.3cm}|p{1.8cm}|}
\hline 
Properties & Types &verification result on $\mathbb{K}$ semantics &testing result on OS kernel \\
\hline
DF & Std &  $\times$ &  \multicolumn{1}{|c|}{ $\surd$ } \\
\hline
ME & Std &   $\centering{\surd}$ &  \multicolumn{1}{|c|}{ $\surd$ } \\
\hline
PIF & Std &  $\surd$ & \multicolumn{1}{|c|}{ $\surd$ }\\
\hline
SF  & Std & $\surd$ &  \multicolumn{1}{|c|}{ $\surd$ }\\
\hline
PE & App & $\surd$ &  \multicolumn{1}{|c|}{ $\surd$ }\\
\hline
MAF & App & $\times$ &  \multicolumn{1}{|c|}{ $\surd$ }  \\
\hline
\end{tabular}
\begin{tablenotes}
        \footnotesize
        \item[1] Std: properties extracted from the OSEK/VDX standard.        
        \item[2] App: properties extracted from EMS .
      \end{tablenotes}
    \end{threeparttable}   
    \end{center}
   \caption{ \label{tab 1} Formal verification and test result}
  \end{table}

\section{Related Work}
\par Some efforts have been done to formally study the OSEK/VDX standard. For example, in work \cite{modelchecking} the standard is partially formalised in Promela and verified by model checker Spin. Because the operating system kernel needs some environment models (applications) to run, they generate a bounded number of environment models for formal analysis. However, they do not consider alarm mechanism in their work. 
In work \cite{ORIENTAIS} and \cite{CSP}, a concrete automobile operating system based on the OSEK/VDX standard is formally specified in CSP and verified by model checker PAT. Their formalisations are built in the code level, and they mainly focus on analysing the operating system instead of the standard.  Another work \cite{yes} presents an approach to verifying the OSEK/VDX applications based on bounded model checking. But they do not take real time features of the system into account. All of the above works are based on model checking technique, which is also supported by  $\mathbb{K}$  framework. 

\par As for the formal analysis of real time features in automobile RTOS, the work \cite{maude} presents an approach to formalising and verifying the schedule table of an AUTOSAR-based RTOS  using Real-Time Maude \cite{realtimemaude}. AUTOSAR (AUTomotive Open System Architecture) \cite{AUTOSAR} is the latest automobile standard based on OSEK/VDX. Real-Time Maude is an extension of Maude, which is also based on rewriting logic like  $\mathbb{K}$ but supports formalisation of real time features. This is the first work on analysing real time features on automobile systems to our knowledge. Inspired by this work, we achieve the formalisation of real time in $\mathbb{K}$. Compared with Real-Time Maude, $\mathbb{K}$ is more scalable to describe complicated and large systems.
For instance, there are several large-scale semantics having been defined in  $\mathbb{K}$, like the semantics of C \cite{KC} and Java \cite{Kjava}.

\par As for the conformance verification of the OSEK/VDX RTOSs,  the work \cite{choi2013} proposes a method for systematically constructing test cases by introducing a constraint specification language, called OSEK\_CSL. They construct test cases automatically using model checker NuSMV. Similarly, the work \cite{spintest} presents an approach to generating test cases by formalising the test requirement using Promela and Z-notation. Together with the formal models, the test cases can be generated by recording the search path in a tool called TGT (test case generation tool). However, both of these two works on conformance checking still relies on traditional testing techniques and mainly focus on the test of system services.

\section{Conclusion}

\par This paper presented an executable formal semantics of the OSEK/VDX standard. The formalisation includes most of the key components in the standard, such as task scheduling, event mechanism, resource management and in particular alarm mechanism. We reported some ambiguities of the standard which are found in the process of formalisation. We then defined a formal definition to these ambiguous descriptions after systematical analysis.
Based on the formal semantics, we proposed an approach to verifying the conformance to the OSEK/VDX standard of an automobile operating system. We finally conducted a case study with an industrial automobile RTOS, which shows the effectiveness of the proposed approach.

\par Compared with the work in \cite{OSEK-K}, the formal semantics of OSEK/VDX RTOS presented in this paper becomes more completed by including the real time features and clarifying some ambiguities. However, some problems mentioned in Section 4.2 remain challenging issues and should be solved before the proposed approach is applied to industry. For example, the interrupt is an essential mechanism of the OSEK/VDX standard. But interrupt may lead to state explosion during formal verification because it can occur almost at any time. In addition, we only considered the formal semantics of system services in the implementation of tasks. We also need to formalise the semantics of other statements in order to verify the properties that are specific to the tasks. Thanks to the $\mathbb{K}$ semantics of C, it can be achieved by integrating these two semantics. However, it will introduce a new challenge, i.e. how to measure the execution time of each statement in a reasonable manner,  which is one piece of our future work.

\par Another future work is to take the environment into consideration. The automobile operating system usually needs to interact with environment. For example, the system may need the engine temperature and the fuel load as parameters, which vary occasionally. Without knowing their values, the formal semantics can not be executed. One possible solution is symbolic execution, which treats variables with uncertain values as symbols \cite{symbolicexe}. Fortunately, symbolic execution is also supported in $\mathbb{K}$ framework \cite{symbolic}. 
\par A complete formal semantics of the OSEK/VDX standard is an essential basis for conformance verification. Another necessity is that we need to acquire all the applications and their execution results in an automobile operating system. To achieve it we should gain insight into not only the operating system kernel and standard but the implementation of applications of the automobile system.

%\input{references}

%\section{Introduction}
%
%The text of the paper begins here.
%
%\appendix
%\section{Appendix Title}
%
%This is the text of the appendix, if you need one.
%
%\acks
%
%Acknowledgments, if needed.

% We recommend abbrvnat bibliography style.
%\bibliographystyle{IEEEtran}
%
\bibliographystyle{abbrvnat}
%\bibliography{references}
% The bibliography should be embedded for final submission.
%%\input{reference}

\end{document}